\documentclass[sigconf]{acmart}

\pdfoutput=1 

\usepackage[utf8]{inputenc}
\usepackage[most]{tcolorbox}

\usepackage{booktabs}

\usepackage{graphicx}
\usepackage{multirow}
\usepackage{setspace}
\usepackage{amsmath}
\usepackage{amsfonts} %
\usepackage{calligra} %
\usepackage{mathtools}
\usepackage{amsthm} 
\usepackage{enumitem}

\usepackage{pgfplots}
\usepackage{pgfplotstable}
\usepgfplotslibrary{statistics}

\makeatletter
\let\MYcaption\@makecaption
\makeatother
\usepackage[font=footnotesize,labelformat=simple]{subcaption}
\makeatletter
\let\@makecaption\MYcaption
\makeatother

\usepackage{comment}
\usepackage{listings}

\usepackage{circledsteps}

\usetikzlibrary{patterns}

\usepackage{blindtext}

\usepackage{acronym}

\usepackage{xcolor, colortbl}

\usepackage{float}
\usepackage{textcomp}

\newcommand{\hammond}[1]{{\color{blue} Hammond: #1}}

\usepackage{algorithm}
\usepackage{algorithmic}

\usepackage{url}

\usepackage[all=normal,paragraphs=tight, floats=tight,indent=tight]{savetrees}
\usepackage{pifont}
\usepackage{framed}
\usepackage{soul}
\usepackage{csquotes}
\usepackage{multirow}
\usepackage{array}
\newcolumntype{L}[1]{>{\raggedright\let\newline\\\arraybackslash\hspace{0pt}}m{#1}}
\newcolumntype{C}[1]{>{\centering\let\newline\\\arraybackslash\hspace{0pt}}m{#1}}
\newcolumntype{R}[1]{>{\raggedleft\let\newline\\\arraybackslash\hspace{0pt}}m{#1}}
\newcolumntype{H}{>{\collectcell\lstinline}l<{\endcollectcell}}
\usepackage{url}
\MakeOuterQuote{"}

\input{section/preamble/preamblelistingsconf}
\acrodef{CPS}{Cyber-Physical System}
\acrodef{IoT}{Internet of Things}
\acrodef{HDL}{Hardware Description Language}
\acrodef{CAD}{Computer-Aided Design}
\acrodef{EDA}{Electronic Design Automation}
\acrodef{HPC}{High-Performance Computing}
\acrodef{DL}{deep learning}
\acrodef{ML}{machine learning}
\acrodef{NLP}{natural language processing}
\acrodef{IC}{Integrated Circuit}
\acrodef{CWE}[CWE]{Common Weakness Enumeration}
\acrodef{CVE}[CVE]{Common Vulnerabilities and Exposures}
\acrodef{LLM}[LLM]{large language model}
\acrodef{NMT}[NMT]{neural machine translation}
\newcommand{\ignore}[1]{{}}

\newcommand{\squishlist}{
	\begin{list}{$\bullet$}
		{ \setlength{\itemsep}{0pt}
			\setlength{\parsep}{1pt}
			\setlength{\topsep}{1pt}
			\setlength{\partopsep}{0pt}
			\setlength{\leftmargin}{0.9em}
			\setlength{\labelwidth}{1.5em}
			\setlength{\labelsep}{0.4em} } }
	\newcommand{\squishend}{
	\end{list}  }

\definecolor{graphFirst}{RGB}{2,136,209} 
\definecolor{graphSecond}{RGB}{211,47,47} 
\definecolor{graphThird}{RGB}{245,124,0} 
\definecolor{graphFourth}{RGB}{56,142,60} 
\definecolor{graphFifth}{RGB}{81,45,168} 
\definecolor{graphSixth}{RGB}{69,90,100} 
\definecolor{graphSeventh}{RGB}{251,192,45} 
\definecolor{backgroundSecond}{RGB}{239,154,154} 
\definecolor{backgroundThird}{RGB}{255,204,128} 
\definecolor{backgroundFourth}{RGB}{165,214,167} 
\definecolor{backgroundFifth}{RGB}{179,157,219} 
\definecolor{backgroundSixth}{RGB}{176,190,197} 
\definecolor{backgroundSeventh}{RGB}{255,245,157} 

\copyrightyear{2023}
\acmYear{2023}
\setcopyright{none}
\renewcommand\footnotetextcopyrightpermission[1]{}


\begin{document}

\fancyhead{}

\thispagestyle{plain}
\pagestyle{plain}


\title{Fixing Hardware Security Bugs with Large Language Models}

\author{Baleegh Ahmad}
\affiliation{%
  \institution{New York University\\ba1283@nyu.edu}
  \country{}
}

\author{Shailja Thakur}
\affiliation{%
  \institution{New York University\\st4920@nyu.edu}
    \country{}
}

\author{Benjamin Tan}
\affiliation{%
  \institution{University of Calgary\\benjamin.tan1@ucalgary.ca}
    \country{}
}

\author{Ramesh Karri}
\affiliation{%
  \institution{New York University\\rkarri@nyu.edu}
    \country{}
}

\author{Hammond Pearce}
\affiliation{%
  \institution{New York University\\hammond.pearce@nyu.edu}
    \country{}
}

\begin{abstract}

Novel AI-based code-writing Large Language Models (LLMs) such as OpenAI's Codex have demonstrated capabilities in many coding-adjacent domains. 
In this work we consider how LLMs maybe leveraged to automatically repair security-relevant bugs present in hardware designs.
We focus on bug repair in code written in the Hardware Description Language Verilog.
For this study we build a corpus of domain-representative hardware security bugs. We then design and implement a framework to quantitatively evaluate the performance of any LLM tasked with fixing the specified bugs. The framework supports design space exploration of prompts (i.e., prompt engineering) and  identifying the best parameters for the LLM. We show that an ensemble of LLMs can repair all ten of our benchmarks. This ensemble outperforms the state-of-the-art Cirfix hardware bug repair tool on its own suite of bugs.
These results show that LLMs can repair hardware security bugs and the framework is an important step towards the ultimate goal of an automated end-to-end bug repair framework.

\end{abstract}

\maketitle

\section{Introduction\label{sec:intro}}

\begin{figure}[t]
    \centering
    \includegraphics[width=\linewidth]{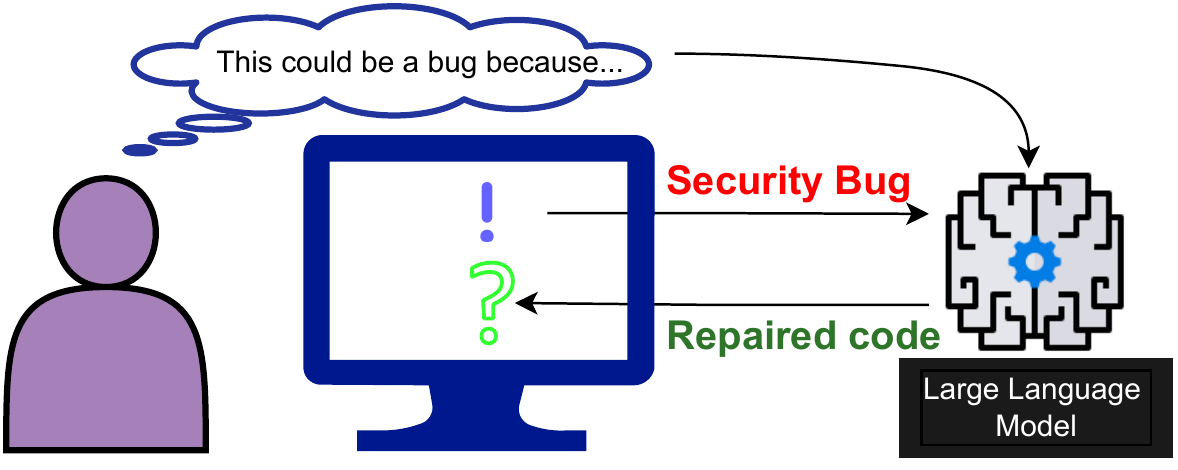}
    \caption{LLMs can suggest repairs to designers.}
    \label{fig:introduction-repair}
\end{figure}

`Bugs' are inevitable when writing large quantities of code.
Fixing them is laborious: automated tools are thus designed and employed to both identify bugs and then patch and repair them~\cite{bader_getafix_2019, goues_automated_2019}. 
While considerable effort has explored software repair, for Hardware Design Languages (HDLs), the state of the art is less mature.

In this study, we focus on repairing security-relevant hardware bugs. While linters \cite{vclint_synopsys_2022, jasperlint_jasper_2022} and formal verification tools \cite{noauthor_vc_2022, cadence_jasper_2022} cover a large proportion of functional bugs, fewer tools cover hardware security bugs. Although formal verification tools like Synopsys FSV can be used for security verification in the design process, they have limited success \cite{dessouky_hardfails_2019}. 
Unlike software bugs, security bugs in hardware are more problematic because they cannot be patched once the chip is fabricated; this is especially concerning as hardware is typically the root of trust for a system~\cite{rostami_primer_2014}. 
With the ever-growing complexity of modern processors, software-exploitable hardware bugs are becoming common and pernicious \cite{lipp_meltdown_2018,kocher_spectre_2019}. This has resulted in the exploration of many techniques such as fuzzing \cite{trippel_fuzzing_2021, tyagi_thehuzz_2022}, information flow tracking \cite{ardeshiricham_register_2017,wu_fault_2022,nahiyan_script_2020}, unique program execution checking \cite{fadiheh_processor_2019} and static analysis \cite{bidmeshki_hunting_2021,ahmad_dont_2022}. However, very few techniques address the automated repair of hardware bugs. The recently proposed Cirfix \cite{ahmad_cirfix_2022} develops automatic repair of functional hardware bugs and, to the best of our knowledge, is the only relevant effort in this context thus far.
Further efforts need to be made to support the automated repair of functional and security bugs in hardware. 

Large Language Models (LLMs) are neural networks trained over millions of lines of text and code \cite{chen_evaluating_2021}. LLMs that are fine-tuned over open-source code repositories can generate code, where a user ``prompts'' the model with some text (e.g., code and comments) to guide the code generation. 
In contrast to previously proposed code repair techniques that involve mutation, repeated checks against an ``oracle,''  or source code templates, we propose that an LLM trained on code and natural language could potentially generate fixes, given an appropriate prompt. 
As LLMs are exposed to a wide variety of code examples during training, they should be able to assist designers in fixing bugs in different types of hardware designs and styles, with natural language guidance. 
In prior work~\cite{thakur_benchmarking_2022, pearce_dave_2020}, LLMs have been used to generate functional Verilog code.
Machine learning-based techniques such as Neural Machine Translation \cite{tufano_empirical_2019} and pre-trained transformers \cite{drain_generating_2021} are explored in the software domain for bug fixes. Pearce et al.~\cite{pearce_examining_2022} use this approach to repair two scenarios of security weaknesses in Verilog code.

\textbf{Thus, in this work, we investigate the use of LLMs to generate repairs for hardware security bugs.} We study the performance of OpenAI Codex and CodeGen LLMs on instances of hardware security bugs. We offer insights into how best to use LLMs for successful repairs. 
An RTL designer can spot a security weakness and the LLM can help to find a fix as shown in Figure \ref{fig:introduction-repair}.
Our contributions are as follows:
\begin{itemize}
    \item Curating a benchmark of hardware security bugs and their corresponding designs. These are open-sourced at~\cite{review_artifacts_2023}.
    \item Automated framework for using LLMs to generate repairs and evaluate them. We make the framework and artifacts produced in this study available~\cite{review_artifacts_2023}.
    \item Automated end-to-end solution to detect, repair and evaluate repairs for certain bugs utilizing static analysis scanners from prior related work~\cite{ahmad_dont_2022}.
    \item Exploration of different LLMs and their parameters to suggest how best to use LLMs in hardware bug repair. These are posed as research questions answered in Section \ref{sec:Results}.
\end{itemize}


\section{Background and Related Work} \label{sec:background} 

Our work borrows ideas from software domain and applies them to the area of hardware design. Since this is not very common, in this section we present some over-arching concepts that better help understand our implementation.

\subsection{Code Repair}
Software code repair techniques continue to evolve (interested readers can see the living review by Monperrus~\cite{monperrus_living_2018}, which contains an ever-growing list of automated repair tools and techniques). 
Generally, techniques try to fix errors through the use of program mutations and repair templates paired with tests to validate any changes~\cite{wang_loopfix_2019,wong_varfix_2021,le_refixar_2021}. 
Feedback loops are constructed with a reference implementation to guide the repair process~\cite{singh_automated_2013,lu_fapr_2021}. 
Other domain-specific tools may also be built to deal with particular areas like build scripts, web, software models, etc. 

Security bugs are critical bug types that can lead to vulnerable systems.
They can be more difficult to detect and repair than functional bugs, which can be detected by classical testing. 
Proving the presence or absence of a security bug is challenging.
This has led to more `creative' kinds of bug repair, including AI-based machine-learning techniques such as neural transfer learning~\cite{chen_neural_2023} and example-based approaches~\cite{zhang_example-based_2022,ma_vurle_2017}. 
ML-based approaches involve memorization and generalization capabilities of neural networks, allowing a greater ability to suggest repairs for ``unseen'' code. The example-based approaches start off with a dataset consisting of pairs of bugs and their repairs. Then, matching algorithms are applied to spot the best repair candidate from the dataset.
Efforts in repair are also explored in other domains like recompilable decompiled code \cite{reiter_automatically_2022}. 

For digital hardware design, the recently proposed CirFix \cite{ahmad_cirfix_2022} attempts to localize bugs in RTL designs and then repair them. The researchers provide the benchmarks they develop for their study, allowing us to apply our methods to compare results. 
While it is the closest work, there are some fundamental differences in the approaches which limit direct comparisons. These differences are described in Table \ref{tbl:cirfix-comparison-table}. 
CirFix performs both localization/identification of the bug and the repair. These two parts can be examined independently, e.g., Tarsel \cite{wu_fault_2022} uses hardware-specific timing information and the program spectrum and captures the changes of executed statements to locate faults effectively. Tarsel outperforms CirFix on CirFix's benchmarks as a fault localizer. 
In our work, we focus on the \textit{repair} aspect. 
Our repair approach has the advantage that an oracle is not needed. While CirFix instruments an oracle to use the correct outputs to guide repairs, LLMs rely on the many examples of RTL code from training to produce a correct version of the buggy code.
We compare our framework's performance with CirFix and discuss it in Section \ref{subsec:results-cirfix}.

\begin{table}[h]
\caption{Comparison with CirFix's approach}
\label{tbl:cirfix-comparison-table}
\begin{tabular}{ll}
\hline
\textbf{CirFix \cite{ahmad_cirfix_2022}}         & \textbf{LLMs (e.g., this study)} \\ \hline \hline
Localization and repair & Repair only (assumes location) \\ \hline
Oracle-guided           & No oracle needed \\ \hline
\begin{tabular}[c]{@{}l@{}}Uses repair templates\\ and operators\end{tabular}   & Uses instructions \\ \hline
Iterative process        & One shot \\ \hline
\end{tabular}
\end{table}

\subsection{Bugs in Register Transfer Level design}

Register Transfer Level (RTL) designs, typically coded in Hardware Description Languages (HDLs) such as Verilog, are high-level behavioral descriptions of hardware circuits specifying how data is transformed, transferred, and stored.
RTL logic features two types of elements, sequential and combinational. Sequential elements (e.g., registers, counters, RAMs) tend to synchronize the circuit according to clock edges and retain values using memory components. Combinational logic (e.g., simple combinations of gates) change their outputs instantaneously according to the inputs.
Whereas software code describes programs that will be executed from beginning to end, RTL specified in HDL describes hardware designs to be implemented. As hardware, components run independently in parallel.

Like software, hardware designs have security bugs.
By definition, RTL is insecure if the security objectives of the circuit are unmet. These may include confidentiality and integrity requirements~\cite{potlapally_hardware_2011}. Confidentiality is violated if data that should not be seen/read under certain conditions is exposed. For example,  improper memory protection allows encryption keys to be read by user code. Integrity is violated if data that should not be modifiable under certain conditions is modifiable. For example,  user code can write into registers that specify the access control policy. Secure computation is a concern, and the synthesis and optimization of secure circuits starts with the description of designs with HDLs~\cite{demmler_automated_2015}. Verisketch~\cite{ardeshiricham_verisketch_2019} defines a synthesis language to implement timing-sensitive information flow properties to generate secure RTL.


\subsection{Static Analysis}
Static Analysis of code involves breaking down the code into its syntactic and lexical elements and exploring this information without simulating/compiling the code. This gives a lot of useful information, primarily in the form of an Abstract Syntax Tree (AST), which contains the variables, signals, operators, keywords, function definitions, parameters, and many other elements. Many tools have utilized static techniques in repair~\cite{berabi_tfix_2021, etemadi_sorald_2022}.
Static analysis is helpful for bug detection and repair as it can be done in the early stages of development. This is particularly beneficial in the hardware domain as once the RTL is synthesized and fabricated into a circuit in silicon, patches are not possible, and the cost of fixing the issue increases exponentially.

\subsection{Common Weakness Enumerations}
MITRE~\cite{the_mitre_corporation_cwe_2022} is a not-for-profit that works with academia and industry to come up with a list of Common Weakness Enumerations (CWEs) that represent categories of vulnerabilities in hardware and software.
A weakness is an element in a digital product's software, firmware, hardware, or service that can be exploited for malicious purposes.
The CWE list provides a general taxonomy and categorization of these elements that allow a common language to be used for discussion. It helps developers and researchers search for the existence of these weaknesses in their designs and compare various tools they use to detect vulnerabilities in their designs and products.
In this work, we address a few CWEs that our designs contain. We identify a CWE that best describes the bug. 
\paragraph{1234: Hardware Internal or Debug Modes Allow Override of Locks} System configuration controls, e.g., memory protection is set after a power reset and then locked to prevent modification. This is done using a lock-bit signal. If the system allows debugging operations and the lock-bit can be overridden in a debug mode, the system configuration controls are not properly protected.
\paragraph{1271: Uninitialized Value on Reset for Registers Holding Security Settings} Security-critical information stored in registers should have a known value when being brought out of reset. If that is not the case, these registers may have unknown values that put the system in a vulnerable state.
\paragraph{1280: Access Control Check Implemented After Asset is Accessed} Access control checks are required in hardware before security-sensitive assets like keys are accessed. If this check is implemented after the access, then the check is clearly useless.
\paragraph{1276: Hardware Child Block Incorrectly Connected to Parent System} Hardware blocks are connected to a parent system that controls their inputs. If an input is incorrectly connected, affecting security attributes like resets while maintaining correct functionality; the integrity of the data of the child block can be violated.
\paragraph{1245: Improper Finite State Machines (FSMs) in Hardware Logic} FSMs are used in hardware to carry out different functionality according to different states. When FSMs are used in modules that control the level of security a system is in, it becomes important that the FSM does not have any undefined states. These undefined states may allow an adversary to carry out functionality that requires higher privileges. An improper FSM can present itself as unreachable states, FSM deadlock, or missing states.

\subsection{Prompt Engineering}
Prompt engineering is crucial to the performance of an LLM. 
Careful prompt engineering outperforms the baseline LLM performances in natural language tasks \cite{zhou_large_2022, strobelt_interactive_2023}. A study exploring the use of Copilot \cite{github_github_2021} to solve CS1 level coding assignments has shown that tweaks to the prompt improve the performance from around 50\% to 60\% \cite{denny_conversing_2022}. Prompt variations are also important in improving the results of text-to-image generation tasks\cite{oppenlaender_taxonomy_2022,liu_design_2022}.
Thus prompt engineering is crucial when using LLMs for code repair.

\section{Designs and Bugs}
\label{sec:designs_and_bugs}

To explore the idea of using LLMs to fix HW security bugs, we first collate and prepare a set of benchmark designs, coming up with ten hardware security bugs from three sources. The sources are CWE descriptions on the MITRE website \cite{the_mitre_corporation_cwe_2022}, OpenTitan System-on-Chip (SoC) \cite{noauthor_hardware_2019} and the Hack@DAC 2021 SoC \cite{hackevent_hackdac21_2022}. Each bug is represented in a design, as described in Table~\ref{tbl:bugs}.

\subsection{ MITRE's CWEs}
We use examples provided in MITRE's hardware design list to come up with simple designs that may represent CWE(s). The bugs and corresponding fixes for this source are shown in \autoref{fig:MITRE-examples}.

\subsubsection{Locked Register} 
This design has a register that is protected by a lock bit. The contents of the register may only be changed when the \texttt{lock\_status} bit is low. In \autoref{fig:locked-register}, a \texttt{debug\_unlocked} signal overrides the \texttt{lock\_status} signal allowing the locked register to be written into even if \texttt{lock\_status} is asserted.

\subsubsection{Lock on Reset} 
This design has a register that holds sensitive information. This register should be assigned a known value on reset. In \autoref{fig:lock-on-reset}, the register \texttt{locked} should have a value assigned under reset, but in this case, there is no reset block.

\subsubsection{Grant Access} 
This design contains a register that should only be modifiable if the \texttt{usr\_id} input is correct. 
In \autoref{fig:grant-access}, the register \texttt{data\_out} is assigned a new value if the \texttt{grant\_access} signal is asserted. This should happen when \texttt{usr\_id} is correct, but since the check happens after writing into \texttt{data\_out} in blocking assignments, \texttt{data\_out} may be modified when the \texttt{usr\_id} is incorrect.

\subsubsection{Trustzone Peripheral}
This design contains a peripheral instantiated in an SoC. To distinguish between trusted and untrusted entities, a signal is used to assign the security level of the peripheral. This is also described as a privilege bit used in Arm TrustZone to define the security level of all connected IPs. In \autoref{fig:tz_peripheral}, the security level of the instantiated peripheral is grounded to zero, which could lead to incorrect privilege escalation of all input data.

\begin{figure}[h]
\centering
\begin{subfigure}[b]{0.95\linewidth}
\begin{lstlisting}[language=verilog, linebackgroundcolor={\ifnum\value{lstnumber}>7
                \ifnum\value{lstnumber}<9
                    \color{pink}
                \fi
            \fi
\ifnum\value{lstnumber}>8
                \ifnum\value{lstnumber}<10
                    \color{green}
                \fi
            \fi}]
module locked_register ( input [15:0] Data_in, 
input clk, resetn, write, lock_status, debug_unlocked, 
output reg [15:0] Data_out );
always @(posedge clk or negedge resetn) begin
    if (~resetn) begin
        Data_out <= 16'h0000;
    end
    else if (write&(~lock_status|debug_unlocked)) begin
    else if (write&~lock_status) begin
        Data_out <= Data_in;
    end
    else if (~write) begin
        Data_out <= Data_out;
    end
end
endmodule 
\end{lstlisting}
\vspace{-3mm}
\caption{Locked Register: Bug - debug signal overrides lock status signal. Fix- remove debug signal in condition.}
\label{fig:locked-register}
\end{subfigure}

\begin{subfigure}[b]{0.95\linewidth}
\begin{lstlisting}[language=verilog, linebackgroundcolor={\ifnum\value{lstnumber}>4
                \ifnum\value{lstnumber}<7
                    \color{pink}
                \fi
            \fi 
\ifnum\value{lstnumber}>6
                \ifnum\value{lstnumber}<10
                    \color{green}
                \fi
            \fi }]
module lock_on_reset ( 
input wire clk, resetn, unlock, d, 
output reg locked );
always @(posedge clk or negedge resetn) begin
    if(unlock) locked <= d;
    else locked <= locked;
    if (~resetn) locked <= 0;
    else if(unlock) locked <= d;
    else locked <= locked;
end
endmodule
\end{lstlisting}
\vspace{-3mm}
\caption{Lock on reset: Bug- register locked is not assigned a value under a reset condition. Fix- locked register is assigned 0 at reset.
}
\label{fig:lock-on-reset}
\end{subfigure}

\begin{subfigure}[b]{0.95\linewidth}
\begin{lstlisting}[language=verilog, linebackgroundcolor={\ifnum\value{lstnumber}>10
                \ifnum\value{lstnumber}<13
                    \color{pink}
                \fi
            \fi 
\ifnum\value{lstnumber}>12
                \ifnum\value{lstnumber}<15
                    \color{green}
                \fi
            \fi             }]
module user_grant_access(data_out, usr_id, data_in, clk, rst_n);
output reg [7:0] data_out;
input wire [2:0] usr_id;
input wire [7:0] data_in;
input wire clk, rst_n;
reg grant_access;
always @ (posedge clk or negedge rst_n)
begin
    if (!rst_n) data_out = 0;
    else begin
        data_out = (grant_access) ? data_in : data_out;
        grant_access = (usr_id == 3'h4) ? 1'b1 : 1'b0;
        grant_access = (usr_id == 3'h4) ? 1'b1 : 1'b0;
        data_out = (grant_access) ? data_in : data_out;
    end
end
endmodule
\end{lstlisting}
\vspace{-3mm}
\caption{Grant access: Bug- \texttt{grant\_access} signal is used before it is assigned a value. Fix- \texttt{grant\_access} signal is used after it is assigned a value.}
\label{fig:grant-access}
\end{subfigure}

\begin{subfigure}[b]{0.95\linewidth}
\begin{lstlisting}[language=verilog, linebackgroundcolor={\ifnum\value{lstnumber}>6
                \ifnum\value{lstnumber}<8
                    \color{pink}
                \fi
            \fi
\ifnum\value{lstnumber}>7
                \ifnum\value{lstnumber}<9
                    \color{green}
                \fi
            \fi
            }]
module soc(clk, rst_n, rdata, rdata_security_level, data_out);
input clk, rst_n, rdata_security_level;
input [31:0] rdata;
output [31:0]  data_out;
    tz_peripheral u_tz_peripheral(
    .clk(clk), .rst_n(rst_n), .data_in(rdata),
    .data_in_security_level(1'b0),
    .data_in_security_level(rdata_security_level),
    .data_out(data_out)  );
endmodule
\end{lstlisting}
\vspace{-3mm}
\caption{TZ peripheral: Bug- security level to peripheral is incorrectly grounded. Fix- security level for data is correctly assigned to parent signal.}
\label{fig:tz_peripheral}
\end{subfigure}

\vspace{-3.5mm}
\caption{MITRE CWE bugs and their corresponding repairs. The repair (green) replaces the bug (red) for a successful fix.}
\label{fig:MITRE-examples}
\end{figure}

\subsection{Google's OpenTitan} 
OpenTitan is an open-source project designed to provide a silicon root of trust. It contains implementations of security measures that make the SoC secure. We inject bugs by tweaking the RTL of these security measures in different modules. The bugs and their corresponding fixes for this source are shown in \autoref{fig:Opentitan-examples}.

\subsubsection{ROM Control} 
This design contains a module that acts as an interface between the ROM and the system bus. The ROM has scrambled contents, and the controller descrambles the content for memory requests. We target the \texttt{COMPARE.CTRL\_FLOW.CONSISTENCY} security measure in the \texttt{rom\_ctrl\_compare} module. A part of this measure is that the \texttt{start\_i} signal should only be asserted in the \texttt{Waiting} state, otherwise, an alert signal is asserted. In \autoref{fig:rom-control}, because of our induced bug, the alert signal is incorrectly asserted when \texttt{start\_i} is high in any state other than \texttt{Waiting}. 

\subsubsection{OTP Control}
This is a one-time programmable memory controller that provides the programmability  for the device's life cycle. It ensures that the correct life cycle transitions are implemented as the entity of the SoC changes among the 4 -- Silicon Creator, Silicon Owner, Application Provider, and the End User. 
We target the \texttt{LCI.FSM.LOCAL\_ESC} security measure in the \texttt{otp\_ctrl\_lci} module. A part of this measure is that the FSM jumps to an error state if the escalation signal is asserted. In \autoref{fig:otp-control}, no error is raised in such a case because of our induced bug.

\subsubsection{Keymanager KMAC} 
This design carries out the Keccak Message Authentication Code (KMAC) and Secure Hashing Algorithm 3 (SHA3) functionality. It is responsible for checking the integrity of the incoming message with the signature produced from the same secret key.
We target the \texttt{KMAC\_IF\_DONE.CTRL.CONSISTENCY} security measure in the \texttt{keymgr\_kmac\_if} module. A part of this measure is that the kmac done signal should not be asserted outside the accepted window, i.e., when the FSM is in the done state. In \autoref{fig:kmac}, because of our induced bug, the kmac done signal is incorrectly asserted in the transmission state \texttt{StTx}.


\begin{figure}
\centering

\begin{subfigure}[b]{0.95\linewidth}
\begin{lstlisting}[language=verilog, linebackgroundcolor={\ifnum\value{lstnumber}>3
                \ifnum\value{lstnumber}<5
                    \color{pink}
                \fi
            \fi 
\ifnum\value{lstnumber}>4
                \ifnum\value{lstnumber}<6
                    \color{green}
                \fi
            \fi }]
// start_i should only be signalled when we're in the Waiting state
// SEC_CM: COMPARE.CTRL_FLOW.CONSISTENCY
logic start_alert;
assign start_alert = start_i && (state_q != Done);
assign start_alert = start_i && (state_q != Waiting);
\end{lstlisting}
\vspace{-3mm}
\caption{ROM Control: Bug- alert asserted when start is high in any state other than Done. Fix- alert asserted when start is high in any state other than Waiting.
}
\label{fig:rom-control}
\end{subfigure}

\begin{subfigure}[b]{0.95\linewidth}
\begin{lstlisting}[language=verilog, linebackgroundcolor={\ifnum\value{lstnumber}>2
                \ifnum\value{lstnumber}<4
                    \color{pink}
                \fi
            \fi 
\ifnum\value{lstnumber}>3
                \ifnum\value{lstnumber}<5
                    \color{green}
                \fi
            \fi             }]
if (escalate_en_i != lc_ctrl_pkg::Off || cnt_err) begin
  state_d = ErrorSt;
  
  fsm_err_o = 1'b1;
  if (error_q == NoError) begin
    error_d = FsmStateError;
  end
end
\end{lstlisting}
\vspace{-3mm}
\caption{OTP Control: Bug- alert is not raised when escalation signal is high. Fix- fsm alert signal is asserted appropriately.}
\label{fig:otp-control}
\end{subfigure}

\begin{subfigure}[b]{0.95\linewidth}
\begin{lstlisting}[language=verilog, linebackgroundcolor={\ifnum\value{lstnumber}>6
                \ifnum\value{lstnumber}<8
                    \color{pink}
                \fi
            \fi
\ifnum\value{lstnumber}>7
                \ifnum\value{lstnumber}<9
                    \color{green}
                \fi
            \fi
            }]
StTx: begin
    valid = 1'b1;
    strb = {IfBytes{1'b1}};
    // transaction accepted
    if (kmac_data_i.ready) begin
      cnt_en = 1'b1;
      kmac_done_vld = 1'b1;
    
      // second to last beat
      if (cnt == CntWidth'(1'b1)) begin
        state_d = StTxLast;
      end
end
\end{lstlisting}
\vspace{-3mm}
\caption{Keymanager KMAC: Bug- kmac done signal is prematurely asserted. Fix- do not assert done signal here.}
\label{fig:kmac}
\end{subfigure}

\vspace{-3.5mm}
\caption{OpenTitan bugs and their corresponding repairs. The repair (green) replaces the bug (red) for a successful fix.}
\label{fig:Opentitan-examples}
\end{figure}

\subsection{Hack@DAC-21}
Hack@DAC-21 examples are bugs in the hardware designs for Hack@DAC 2021 CTF competition. Hack@DAC is a hackathon for finding vulnerabilities at the RTL level for a reasonably complex System-on-Chip (SoC). The bugs and their corresponding fixes for this source are shown in \autoref{fig:hackdac-examples}.

\subsubsection{Csr regfile} 
This design contains a module that carries out changes in control and status registers according to the system's state. This includes changes in privilege levels, incoming interrupts, virtualization, and cache support. We consider the module's functionality pertaining to the stalling of the core in the case of receiving an interrupt and/or debug request. In \autoref{fig:csr-regfile}, the debug signal overrides interrupt signals.

\subsubsection{DMA} 
This design contains the Direct Memory Access module common to all blocks. It uses the memory address as input and performs read or write operations according to the Physical Memory Protection (PMP) configuration. We consider the PMP access mechanism as the relevant security implementation. In \autoref{fig:dma}, the pmp register is not assigned any value on reset.

\subsubsection{AES 2 Interface}
This design instantiates the Advanced Encryption Standard (AES) module and outputs the cipher text to the system. It uses an FSM to interact with the AES (initialize, reset, and checking valid output). In \autoref{fig:aes2-interface}, the case statement has neither enough cases nor a default statement.

\begin{figure}
\centering

\begin{subfigure}[b]{0.95\linewidth}
\begin{lstlisting}[language=verilog, linebackgroundcolor={\ifnum\value{lstnumber}>4
                \ifnum\value{lstnumber}<6
                    \color{pink}
                \fi
            \fi 
\ifnum\value{lstnumber}>5
                \ifnum\value{lstnumber}<7
                    \color{green}
                \fi
            \fi }]
// Wait for Interrupt
always_comb begin : wfi_ctrl
    // wait for interrupt register
    wfi_d = wfi_q;
    if (|mip_q || debug_req_i || irq_i[1]) begin
    if (|mip_q || irq_i[1]) begin
        wfi_d = 1'b0;
    end else if (!debug_mode_q && csr_op_i == WFI && !ex_i.valid) begin
        wfi_d = 1'b1;
    end
end
\end{lstlisting}
\vspace{-3mm}
\caption{Csr regfile: Bug- debug signal overrides interrupt signals. Fix- remove debug signal in condition.
}
\label{fig:csr-regfile}
\end{subfigure}

\begin{subfigure}[b]{0.95\linewidth}
\begin{lstlisting}[language=verilog, linebackgroundcolor={\ifnum\value{lstnumber}>1
                \ifnum\value{lstnumber}<3
                    \color{pink}
                \fi
            \fi 
\ifnum\value{lstnumber}>2
                \ifnum\value{lstnumber}<7
                    \color{green}
                \fi
            \fi             }]
riscv::pmp_access_t pmp_access_type_reg, pmp_access_type_new; // riscv::ACCESS_WRITE or riscv::ACCESS_READ
reg pmp_access_type_en; 
reg pmp_access_type_en; 
always @ (posedge clk_i or negedge rst_ni) begin
    if (!rst_ni) begin
        pmp_access_type_en  <= 0;
\end{lstlisting}
\vspace{-3mm}
\caption{DMA: Bug- pmp enable register is not assigned a value on reset. Fix- pmp enable register is assigned 0 on reset.}
\label{fig:dma}
\end{subfigure}

\begin{subfigure}[b]{0.95\linewidth}
\begin{lstlisting}[language=verilog, linebackgroundcolor={\ifnum\value{lstnumber}>5
                \ifnum\value{lstnumber}<7
                    \color{pink}
                \fi
            \fi
\ifnum\value{lstnumber}>6
                \ifnum\value{lstnumber}<10
                    \color{green}
                \fi
            \fi
            }]
s15: begin
        Out_data_final <= Out_data;
        ct_valid_out <= 1'b1;
        state <= s0;
    end
    
    default: begin
             state <= s0;
    end
endcase
\end{lstlisting}
\vspace{-3mm}
\caption{AES2 Interface: Bug- Incomplete case statements. Fix- add default case.}
\label{fig:aes2-interface}
\end{subfigure}

\vspace{-3.5mm}
\caption{Hack@DAC-21 bugs and their corresponding repairs. The repair (green) replaces the bug (red) for a successful fix.}
\label{fig:hackdac-examples}
\end{figure}

\begin{table*}[h]
\caption{Bugs Overview. We assign a CWE to each bug and give a description of the design.}
\label{tbl:bugs}

\resizebox{\textwidth}{!}{
\begin{tabular}{lllll}
\hline
Bug & Design & CWE  & Source  & Description  \\ \hline

1   & \begin{tabular}[c]{@{}l@{}}Locked \\ Register\end{tabular} & 1234 & MITRE     & \begin{tabular}[c]{@{}l@{}}This register module supports a lock mode that blocks any writes after lock is set to 1.\\ However, it also allows override of the lock protection when scan\_mode or debug\_unlocked modes are active.\end{tabular}                                                                                                                   \\ \hline
2   & \begin{tabular}[c]{@{}l@{}}Lock on \\ Reset\end{tabular} & 1271 & MITRE     & \begin{tabular}[c]{@{}l@{}}This register module supports a lock mode that allows writes after unlock is set to 1. The locked register \\ does not have a value assigned on reset and when the circuit is brought out of reset, the state will be unknown. \end{tabular} \\ \hline

3   & \begin{tabular}[c]{@{}l@{}} Grant \\ Access\end{tabular}  & 1280 & MITRE     & \begin{tabular}[c]{@{}l@{}} This module allows register contents to be modified only when correct user id is used.\\
However, the asset is allowed to be modified even before the access control check is complete.\end{tabular}  
\\ \hline

4   & \begin{tabular}[c]{@{}l@{}} Trustzone \\ Peripheral\end{tabular}  & 1276 & MITRE     & \begin{tabular}[c]{@{}l@{}} This module instantiates a peripheral within a SoC using a signal to distinguish between trusted and untrusted\\entities.
However, this signal depicting the security level is incorrectly grounded.\end{tabular}  

\\ \hline
5   & \begin{tabular}[c]{@{}l@{}} ROM \\ Control\end{tabular}    & 1245 & OpenTitan & \begin{tabular}[c]{@{}l@{}} This module contains an FSM where an alert should be triggered if start signal is high in any state other than\\Waiting. However, the state is incorrectly compared to the Done state instead of the Waiting state.\end{tabular}   \\ \hline

6   & \begin{tabular}[c]{@{}l@{}} OTP \\ Control\end{tabular}  & 1245 & OpenTitan & \begin{tabular}[c]{@{}l@{}}The life cycle interface FSM should move into an invalid state upon global escalation via life cycle.\\However, the corresponding error signal for this transition is not asserted when the escalation signal is high.\end{tabular}                                                                  \\ \hline
7   & \begin{tabular}[c]{@{}l@{}}Keymanager\\ KMAC\end{tabular}  & 1245 & OpenTitan & \begin{tabular}[c]{@{}l@{}}This module has an FSM which has a done signal which should only be asserted at the time of completion.\\However, this signal is asserted outside of expected window, i.e., during a transmission state.\end{tabular}    \\ \hline

8   & Csr regfile & 1234 & H@DAC-21  & \begin{tabular}[c]{@{}l@{}}If there is any interrupt pending or an incoming interrupt request is received, the core should be unstalled. \\ In this example, the core is also unstalled if there is a request to enter debug mode.\end{tabular}                \\ \hline

9   & DMA  & 1271 & H@DAC-21  & \begin{tabular}[c]{@{}l@{}}This module has a security sensitive register that controls whether the PMP (Physical Memory Protection)\\register can be written into. This register should be assigned a value on reset but it is not.\end{tabular}   
\\ \hline

10   & \begin{tabular}[c]{@{}l@{}}AES-2 \\ interface\end{tabular} & 1245 & H@DAC-21  & \begin{tabular}[c]{@{}l@{}}The FSM for AES 2 interface has a total of 15 states and does not include a default statement for its 4 bit\\ state variable. This represents an incomplete case statement of an FSM. \end{tabular}                                                                                                                                                                                                                      \\ \hline
\end{tabular}
}

\end{table*}

\section{Experimental Method\label{sec:Methodology}}

To test the capability of LLMs to generate successful repairs, we design experiments that use the designs and bugs detailed in \autoref{sec:designs_and_bugs}. In this section we present our framework that automates the execution of our experiments, starting from the identification of bugs to the evaluation of the repairs.

\subsection{LLM-based Repair Evaluation Framework }
\begin{figure}[t]
    \centering
    \includegraphics[width=\linewidth]{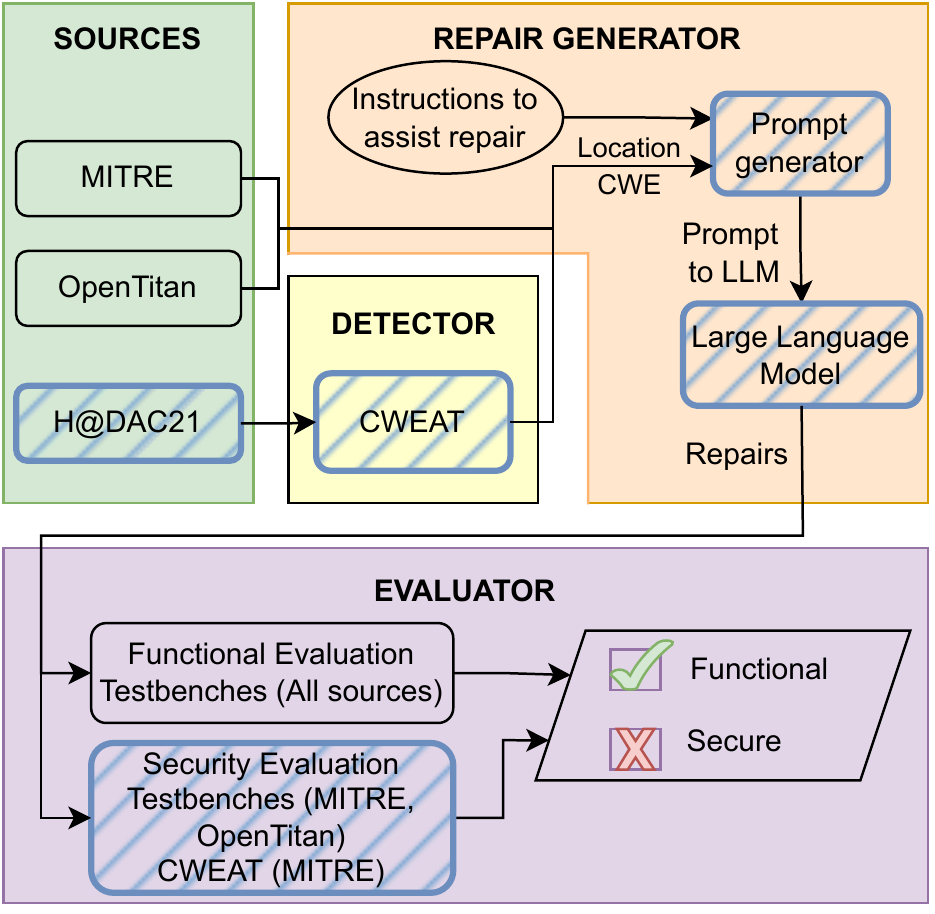}
    \caption{Overview of the framework used in our experiments It is broken down into 4 main components. Sources are the designs containing bugs. Detector localizes the bug (for bugs 8-10). Repair generator contains the LLM which generates the repairs. Evaluator verifies the success of the repair.}
    \label{fig:framework}
\end{figure}

The framework overview for our experiments is shown in \autoref{fig:framework}. It can be broken down into four components, i.e., the \textbf{Sources}, \textbf{Detector}, \textbf{Repair Generator}, and \textbf{Evaluator}. The Sources are discussed in Section \ref{sec:designs_and_bugs}, and the Detector, used for bugs from Hack@ DAC-21, is discussed in Section \ref{subsec:CWEAT}.

\subsubsection{Repair Generator\label{subsubsec:repair-generator}}
This block takes the \textbf{location} and \textbf{CWE} of the bug as the input from the Source or the Detector. For MITRE and OpenTitan, we assume that the location of the bugs, i.e., starting and ending line numbers and the filepath of the buggy file, is known. For Hack@DAC, we run a bug detector tool that gives us the location and relevant CWE of the bugs as its outputs.

For each bug, we develop \textbf{Instructions to assist repair}. These are comments before and after the buggy code that assist the LLMs in generating an appropriate repair for that bug. The \textbf{Prompt generator} combines the code before the bug, buggy code in comments, and instructions to form the \textbf{Prompt to LLM}. This can be worded as `what the LLM sees'. An example of this construction is shown in \autoref{fig:prompt-to-LLM} (a)-(c) for the design Grant Access. The instructions are broken down into Bug Instruction and Fix Instruction. The former describes the nature of the bug and lets the LLM know that the bug follows. The latter follows the bug in comments and instructs the LLM on how to fix the bug. These instructions are varied in different degrees of detail according to the bug as discussed in Section \ref{subsec:instruction-variation}.
The \textbf{Large Language Model} takes the  \textbf{Prompt to LLM} as input and outputs the \textbf{Repairs}. The repairs produced may be correct or incorrect. Some of the repairs generated using the prompt \autoref{fig:prompt-to-LLM-complete} are shown in \autoref{fig:prompt-to-LLM} (d)-(f).

\begin{figure}
\centering

\begin{subfigure}[b]{0.95\linewidth}
\begin{lstlisting}[language=verilog, linebackgroundcolor={\ifnum\value{lstnumber}>9
                \ifnum\value{lstnumber}<12
                    \color{pink}
                \fi
            \fi        }]
module user_grant_access(data_out, usr_id,data_in,clk,rst_n);
output reg [7:0] data_out;
input wire [2:0] usr_id;
input wire [7:0] data_in;
input wire clk, rst_n;
reg grant_access;
always @ (posedge clk or negedge rst_n) begin
    if (!rst_n) data_out = 0;
    else begin
        data_out = (grant_access) ? data_in : data_out;
        grant_access = (usr_id == 3'h4) ? 1'b1 : 1'b0;
    end
end
endmodule
\end{lstlisting}
\vspace{-3mm}
\caption{Original buggy file for bug 3 in design Grant Access.
}
\label{fig:original-file}
\end{subfigure}

\begin{subfigure}[b]{0.95\linewidth}
\begin{lstlisting}[language=verilog, linebackgroundcolor={\ifnum\value{lstnumber}>0
                \ifnum\value{lstnumber}<3
                    \color{yellow}
                \fi
            \fi       
\ifnum\value{lstnumber}>3
                \ifnum\value{lstnumber}<7
                    \color{lightgray}
                \fi
            \fi       }]
// BUG: Access Control Check Implemented after 
// asset is Accessed.

// Ensure that access is granted before data is 
// accessed.
// FIX:
\end{lstlisting}
\vspace{-3mm}
\caption{Instructions used to assist repair utilizing variation c. The Bug instruction is highlighted in yellow and the Fix instruction is highlighted in gray.}
\label{fig:Instructions}
\end{subfigure}

\begin{subfigure}[b]{0.95\linewidth}
\begin{lstlisting}[language=verilog, linebackgroundcolor={\ifnum\value{lstnumber}>10
                \ifnum\value{lstnumber}<13
                    \color{yellow}
                \fi
            \fi
\ifnum\value{lstnumber}>12
                \ifnum\value{lstnumber}<15
                    \color{pink}
                \fi
            \fi
\ifnum\value{lstnumber}>14
                \ifnum\value{lstnumber}<18
                    \color{lightgray}
                \fi
            \fi         }]
module user_grant_access(data_out,usr_id,data_in,clk,rst_n);
output reg [7:0] data_out;
input wire [2:0] usr_id;
input wire [7:0] data_in;
input wire clk, rst_n;
reg grant_access;
always @ (posedge clk or negedge rst_n) begin
    if (!rst_n)
        data_out = 0;
    else begin
// BUG: Access Control Check Implemented after 
// asset is Accessed.
// data_out = (grant_access) ? data_in : data_out;
// grant_access = (usr_id == 3'h4) ? 1'b1 : 1'b0;
// Ensure that access is granted before data is 
// accessed.
// FIX:
\end{lstlisting}
\vspace{-3mm}
\caption{Prompt-to-LLM: What the LLM gets as an input.}
\label{fig:prompt-to-LLM-complete}
\end{subfigure}

\begin{subfigure}[b]{0.95\linewidth}
\begin{lstlisting}[language=verilog]
grant_access = (usr_id == 3'h4) ? 1'b1 : 1'b0;
if (grant_access) data_out = data_in;
\end{lstlisting}
\vspace{-3mm}
\caption{Correct repair produced by \texttt{code-davinci-001} at temp=0.7.}
\label{fig:correct-repair}
\end{subfigure}

\begin{subfigure}[b]{0.95\linewidth}
\begin{lstlisting}[language=verilog]
data_out = (grant_access) ? data_in : data_out;
grant_access = (usr_id == 3'h4) ? 1'b1 : 1'b0;
\end{lstlisting}
\vspace{-3mm}
\caption{Incorrect repair produced by \texttt{code-davinci-001} at temp=0.5. This response passed the functional evaluation but failed the security evaluation.}
\label{fig:incorrect-repair}
\end{subfigure}

\begin{subfigure}[b]{0.95\linewidth}
\begin{lstlisting}[language=verilog]
if (grant_access) data_out = data_in;
\end{lstlisting}
\vspace{-3mm}
\caption{Incorrect repair produced by \texttt{code-davinci-001} at temp=0.7. This response failed the functional and security evaluations.}
\label{fig:functionally-incorrect-repair}
\end{subfigure}

\begin{subfigure}[b]{0.95\linewidth}
\begin{lstlisting}[language=verilog, linebackgroundcolor={\ifnum\value{lstnumber}>10
                \ifnum\value{lstnumber}<13
                    \color{green}
                \fi
            \fi}]
module user_grant_access(data_out,usr_id,data_in,clk,rst_n);
output reg [7:0] data_out;
input wire [2:0] usr_id;
input wire [7:0] data_in;
input wire clk, rst_n;
reg grant_access;
always @ (posedge clk or negedge rst_n) begin
    if (!rst_n)
        data_out = 0;
    else begin
        grant_access = (usr_id == 3'h4) ? 1'b1 : 1'b0;
        if (grant_access) data_out = data_in;
    end
end
endmodule
\end{lstlisting}
\vspace{-3mm}
\caption{Generated repair file using \ref{fig:correct-repair}. This is sent to the Evaluator for evaluation.}
\label{fig:repair-for-evaluation}
\end{subfigure}

\vspace{-3.5mm}
\caption{Prompt to LLM and sample repairs produced for Bug 3 - Grant Access. Sub-figures (a)-(c) show how the bug is combined with instructions to generate the prompt that the LLM gets as one of its inputs. Sub-figures (d)-(f) show some actual repairs generated by an LLM.}
\label{fig:prompt-to-LLM}
\end{figure}

\subsubsection{Evaluator}
This block takes the \textbf{Repairs} generated by the LLM and verifies their correctness by evaluating their functionality and security. A repair is successful if it is both \textbf{functional} and \textbf{secure}. We use ModelSim simulator as a part of Xilinx Vivado 2022.2 to simulate the designs and custom testbenches.

\textbf{Functional Evaluation} is done using custom testbenches we developed in Verilog. These are made for each design and contain tests to check for various input vectors. A failed testbench indicates a failure of at least one test or a syntax error in the design. For MITRE designs, we develop testbenches that cover the design's entire intended functionality. For OpenTitan and Hack@DAC designs, we cover partial functionality for inputs and outputs that pertain to the buggy code. These designs require an additional step of forming the Device Under Test (DUT) before simulation. This entails tracking the files instantiated by the buggy file and the files that need to be analyzed before the buggy file. This list of files is input to the simulator. 

\textbf{Security Evaluation} is done through a combination of custom testbenches (for MITRE and OpenTitan) and CWEAT (for Hack@DAC). For MITRE designs, the tests are designed according to the weaknesses mentioned on the MITRE website for each bug. For OpenTitan, we use the security countermeasures defined in their relevant .hjson files for the peripherals. It is difficult to verify the security countermeasure completely because that requires simulating the entire SoC through the software for Design Verification by OpenTitan. This method is still a work in progress for the OpenTitan team. The countermeasures that can currently be verified completely still require a lot of simulation time. Hence, we develop custom testbenches that verify very specific functionality for the bugs we introduce in the OpenTitan designs. For Hack@DAC, we employ CWEAT for security evaluation; this is discussed in Section \ref{subsec:CWEAT}.

Functional and Security Verification are not always mutually exclusive. There is often an overlap between the two, e.g., for CWE 1271 bugs, the security verification requires both a value on reset for the security-sensitive register and the correct lock mechanism. The latter is also a requirement for correct functionality. In the case of Bug 3, the functional verification is a subset of the security evaluation because the goal of the design is to grant user access under the correct input.

\subsection{End-to-end example with CWEAT \label{subsec:CWEAT} }
We present a demonstrative end-to-end framework for the detection and repair of some CWEs in Verilog. This includes the detection of the bug, the generation of repair using this detection, and the evaluation of the correctness of the repair generated. The elements of this pipeline are represented in hatched blocks in \autoref{fig:framework}.

The \textbf{Detector} used is a static analysis tool that has the capability to detect some weaknesses at the RTL. We use the methods described in \cite{ahmad_dont_2022} to traverse the Abstract Syntax Trees (AST)s generated by the Verific Verilog parser. There is one AST produced per module. Each node of the tree represents a syntactical element of the RTL code with various information about identifiers, types, values and conditions. The ASTs are traversed using keywords and patterns to indicate potential vulnerabilities in CWEs 1234, 1271, and 1245. We ran this tool over the Hack@DAC 2021 SoC and selected three instances, one per CWE, for the purposes of this paper.
We use the same tool for \textbf{security evaluation} of the generated responses. We replace the buggy code with the repaired code in the SoC and run the tool again. If the same bug is picked up, i.e., the same location and CWE, we can determine that the repair is not successful. If that is not the case, we infer that the repair is adequate. 

We envision the use of this (or similar) LLM-infused end-to-end solution by RTL designers as they write Hardware Description Language (HDL) code in the early stages of Hardware Design. CWEAT can highlight the potential weakness to the designer, run it through the LLM to produce repairs, choose the ones that are secure, and present those as suggestions to the designer. 

\subsection{Experimental Parameters}
LLMs have several parameters that can be manipulated to produce responses. We change the prompt (as discussed in Section \ref{subsubsec:repair-generator}) according to the bug and instructions. We also vary the \textbf{Instructions}, \textbf{Temperature} and \textbf{Models} while keeping the \textbf{top\_p}, \textbf{number\_of\_completions (n)} and \textbf{max\_tokens} constant at 1, 20 and 200 respectively. \textbf{top\_p} is an alternative to sampling with temperature, called nucleus sampling, where only results with probability mass of \textbf{top\_p} are considered. \textbf{n} is the number of completions generated by the LLM per request. \textbf{max\_tokens} is the maximum number of tokens that can be generated per completion.

\subsubsection{Instruction Variation \label{subsec:instruction-variation} }
We test five instruction variants to guide the repair of bugs. They are described in Table \ref{tbl:summary-instruction-variations}. Each variation has 2 parts -- \textbf{Bug Instruction} and \textbf{Fix Instruction}. The former describes the nature of the bug and precedes the commented bug. The latter follows the bug in comments and represents guidance to the LLM on how to fix the bug.

Variation \textbf{a} provides no assistance and is the same across all bugs. The \textbf{Bug instruction} is \texttt{"BUG:"} and the \textbf{Fix Instruction} is \texttt{"FIX:"}.
The \textbf{Bug Instruction} for the remaining variations is a description of the nature of the bug. We take inspiration from the MITRE website and cater them according to the CWE they represent. For variation \textbf{e} this description is appended with an example of a `generalized' bug in comments and its fix without comments. This generalization is done through using more common signal names and coding patterns.
The \textbf{Fix Instruction} for \textbf{b} and \textbf{e} is the same as that for \textbf{a}. For \textbf{c}, it is preceded by a `prescriptive' instruction which means that natural language is used to assist the fix. For \textbf{d}, however, it is preceded by a `descriptive' instruction which means that language resembling pseudo-code is used to assist the fix. The components of instruction that change are shown in Table \ref{tbl:details-instruction-variations}.

\begin{table}[]
\caption{Instruction Variations. We develop 5 types to assist repair of bugs. Variation a is the base variation with no assistance. The level of detail/assistance increases from variation a to e.}
\label{tbl:summary-instruction-variations}
\begin{tabular}{cl}
\hline
\begin{tabular}[c]{@{}c@{}}Instruction\\ Variation\end{tabular} & Description                                                                                                           \\ \hline\hline
a                                                               & No Instruction                                                                                                                \\ \hline
b                                                               & Natural language description of bug                                                                                   \\ \hline
c                                                               & \begin{tabular}[c]{@{}l@{}}Natural language description of bug \\ Prescriptive instruction of how to fix\end{tabular} \\ \hline
d                                                               & \begin{tabular}[c]{@{}l@{}}Natural language description of bug\\ Descriptive instruction of how to fix\end{tabular}   \\ \hline
e                                                               & Code examples of bug and fix                                                                                          \\ \hline
\end{tabular}
\end{table}

\begin{table*}
\caption{Details of Instruction Variations and Stop keywords used. The same Bug instruction is used for variations $b$,$c$,$d$, shown in column 2. In case of variation $e$, this Bug instruction (in column 2) is appended by an example of a bug and its repair in comments, shown in column 3. Fix instructions for variations c and d precede the string “FIX:”, shown in columns 4 and 5 respectively. Additional stop keywords that terminate the further generation of tokens by LLMs are shown in column 6.}
\label{tbl:details-instruction-variations}
\resizebox{\textwidth}{!}{
\begin{tabular}{cllllc}
\hline
Bug & 
\begin{tabular}[c]{@{}l@{}}Bug Instruction for \\ variations $b$,$c$,$d$,$e$\end{tabular}& 
\begin{tabular}[c]{@{}l@{}}Bug Instruction appended \\ for variation $e$\end{tabular}& 
\begin{tabular}[c]{@{}l@{}}Fix Instruction \\ for variation $c$\end{tabular} & 
\begin{tabular}[c]{@{}l@{}}Fix Instruction \\ for variation $d$\end{tabular} &
\begin{tabular}[c]{@{}l@{}}Stop \\ keywords \end{tabular}

\\ \hline \hline

1   & \begin{tabular}[c]{@{}l@{}}// BUG: Hardware Internal\\ or Debug Modes Allow\\ Override of Locks.\end{tabular}             & \cellcolor[HTML]{FFFFFF}\begin{tabular}[c]{@{}l@{}}// if (write \& ($\sim$lock | debug)\\ // FIX:\\  if (write \& $\sim$lock)\end{tabular}                                                                              & \cellcolor[HTML]{FFFFFF}\begin{tabular}[c]{@{}l@{}}// Debug or scan signals \\ should not be included\\ in if condition\end{tabular}        & \begin{tabular}[c]{@{}l@{}}// Write data if write\\ signal high and lock\\ signal is low\end{tabular} 
 & `$\backslash n$'
\\ \hline

2   & \begin{tabular}[c]{@{}l@{}}// BUG: Uninitialized Value\\ on Reset for Registers\\ Holding Security Settings.\end{tabular} & \begin{tabular}[c]{@{}l@{}}// if(unlock) q \textless{}= d\_in; else q \textless{}= q;\\ // FIX:\\     if($\sim$resetn) q\textless{}=0; \\     else if(unlock) q \textless{}= d\_in; else q \textless{}= q;\end{tabular} & \begin{tabular}[c]{@{}l@{}}// Ensure that the security\\ sensitive lock register is\\ assigned a value on reset.\end{tabular}               & \begin{tabular}[c]{@{}l@{}}// Assign 0 to register\\ when reset is low\end{tabular}            & -
\\ \hline

3   & \begin{tabular}[c]{@{}l@{}}// BUG: Access Control \\ Check Implemented\\ After Asset is Accessed.\end{tabular}            & \begin{tabular}[c]{@{}l@{}}// d = (access) ? d\_in : d\_out;\\ // access = (id == 2'h2) ? 1'b1 : 1'b0;\\ // FIX:\\     access = (id == 2'h2) ? 1'b1 : 1'b0;\\     d = (access) ? d\_in : d\_out;\end{tabular}           & \begin{tabular}[c]{@{}l@{}}// Ensure that access is\\ granted before data is\\ accessed.\end{tabular}                                       & \begin{tabular}[c]{@{}l@{}}// Assert access when\\ id is correct. Then\\ assign data to register\\ if access is asserted.\end{tabular}
 & end
\\ \hline

4   & \begin{tabular}[c]{@{}l@{}}// BUG: Hardware Child \\ Block Incorrectly Connected\\ to Parent System\end{tabular}          & \begin{tabular}[c]{@{}l@{}}// .in\_security\_level(1'b0),\\ // FIX:\\ .in\_security\_level(data\_security\_level),\end{tabular}                                                                                         & \begin{tabular}[c]{@{}l@{}}// The security level of\\ the child signal should\\ match that of the parent\\ signal\end{tabular}              & \begin{tabular}[c]{@{}l@{}}// assign data security\\ level to input security\\ level\end{tabular}                           & `$\backslash n$'
\\ \hline

5   & \begin{tabular}[c]{@{}l@{}}// BUG: Incorrect Alert\\ Mechanism\end{tabular}                                               & \begin{tabular}[c]{@{}l@{}}// alert = start \&\& (state!=FINISHED);\\ // FIX:\\     alert = start \&\& (state!=IDLE);\end{tabular}                                                                                      & \begin{tabular}[c]{@{}l@{}}// An alert signal should\\ be set if an FSM is\\ instructed to start in a\\ state that is not idle\end{tabular} & \begin{tabular}[c]{@{}l@{}}// Assert alert signal\\ if start signal is\\ asserted and state is\\ not idle\end{tabular}
 & `$\backslash n$'
\\ \hline

6   & \begin{tabular}[c]{@{}l@{}}// BUG: Escalation does\\ not lead to fatal error\end{tabular}                                 & \begin{tabular}[c]{@{}l@{}}// if (escalate\_i != 0 ) begin\\ // state\_d = err\_state;\\ // FIX:\\ if (escalate\_i != 0 ) begin\\       state\_d = err\_state; fsm\_err\_o = 1'b1;\end{tabular}                         & \begin{tabular}[c]{@{}l@{}}// FSM should raise\\ error if system is in\\ escalation\end{tabular}                                            & \begin{tabular}[c]{@{}l@{}}// Assert error when\\ escalation input is\\ high\end{tabular}                                       & end
\\ \hline

7   & \begin{tabular}[c]{@{}l@{}}// BUG: Done signal is\\ asserted prematurely\end{tabular}                                     & \begin{tabular}[c]{@{}l@{}}//      if (ready) begin done\_vld = 1'b1;\\ // FIX:\\       if (ready) begin done\_vld = 1'b0;\end{tabular}                                                                                 & \begin{tabular}[c]{@{}l@{}}// Do not assert done\\ signal in intermediate\\ states\end{tabular}                                             & \begin{tabular}[c]{@{}l@{}}// assign zero to done\\ signal in ready state\end{tabular}           
 & end
\\ \hline

8   & \begin{tabular}[c]{@{}l@{}}// BUG: Hardware Internal\\ or Debug Modes Allow \\ Override of Locks.\end{tabular}            & \begin{tabular}[c]{@{}l@{}}// BUG: Hardware Internal\\ or Debug Modes Allow\\ Override of Locks.\end{tabular} & \begin{tabular}[c]{@{}l@{}}// Debug or scan signals\\ should not be included\\ in if condition\end{tabular}                                 & \begin{tabular}[c]{@{}l@{}}// unstall core when\\ interrupt is high\end{tabular}      & `$\backslash n$'
\\ \hline

9   & \begin{tabular}[c]{@{}l@{}}// BUG: Uninitialized Value\\ on Reset for Registers\\ Holding Security Settings.\end{tabular} & \begin{tabular}[c]{@{}l@{}}// if(unlock) q \textless{}= d\_in; else q \textless{}= q;\\ // FIX:\\     if($\sim$resetn) q\textless{}=0; \\     else if(unlock) q \textless{}= d\_in; else q \textless{}= q;\end{tabular} & \begin{tabular}[c]{@{}l@{}}// Ensure that the security\\ sensitive lock register is\\ assigned a value on reset.\end{tabular}               & \begin{tabular}[c]{@{}l@{}}// Assign 0 to register\\ when reset is low\end{tabular}            & -
\\ \hline

10  & \begin{tabular}[c]{@{}l@{}}// BUG: Incomplete\\ case statement\end{tabular}                                               & \begin{tabular}[c]{@{}l@{}}// endcase\\ // FIX:\\     default: begin\\         state \textless{}= s0; end endcase\end{tabular}                                                                                          & \begin{tabular}[c]{@{}l@{}}// Add a default case\\ statement\end{tabular}                                                                   & \begin{tabular}[c]{@{}l@{}}// Write a default case\\ statement where\\ initial state is\\ assigned to state\end{tabular}               
 & endcase
\\ \hline

\end{tabular}
}
\end{table*}

\subsubsection{Temperature (t)} A higher value means that the LLM takes more risks and yields more creative completions. We use $t \in \{0.1,0.3.0.5,0.7,0.9\}$.
\subsubsection{Models}
We use four LLMs, three of which are made available by OpenAI \cite{openai_openai_2021} and one is an open-source model available through~\cite{nijkamp_codegen_2022}. 
The OpenAI Codex models are derived from GPT-3 and were trained on millions of public GitHub repositories. They can ingest and generate code, and also translate natural language to code. We use/evaluate \texttt{code-davinci-001}, \texttt{code-davinci-002} and \texttt{code-cushman-001} models.
From Hugging Face, we evaluate the model \texttt{CodeGen-16B-multi}, which we refer to as \texttt{CodeGen} in this work. It is an autoregressive language model for program synthesis trained sequentially on The Pile and BigQuery. 



\subsubsection{Number of lines before bug}
Another parameter to consider is in the prompt preparation: the number of lines of existing code given to the LLM. Some files may be too large for the entire code before the bug to be sent to the LLM. We, therefore, select a minimum of 25 and a maximum of 50 lines of code before the bug as part of the prompt. In \autoref{fig:original-file}, this would be lines 1--9 (inclusive). If there are more than 25 lines above the bug, we include enough lines that go up to the beginning of the block the bug is in. This block could be an always block, module, or case statement, etc. 





 



\subsubsection{Stop keywords}
A stop keyword is specified to stop the response of the LLM (i.e., the response is considered finished when the stop keyword is generated by the model). 
They are not included in the response. We developed a strategy that works well with the set of bugs we have. The default stop keyword is \texttt{endmodule}. Additional keywords used are shown in the column Stop keywords in Table \ref{tbl:details-instruction-variations}.







\section{Experimental Results}
\label{sec:Results}

\begin{figure*}[]
    \includegraphics[width=0.95\textwidth]
    {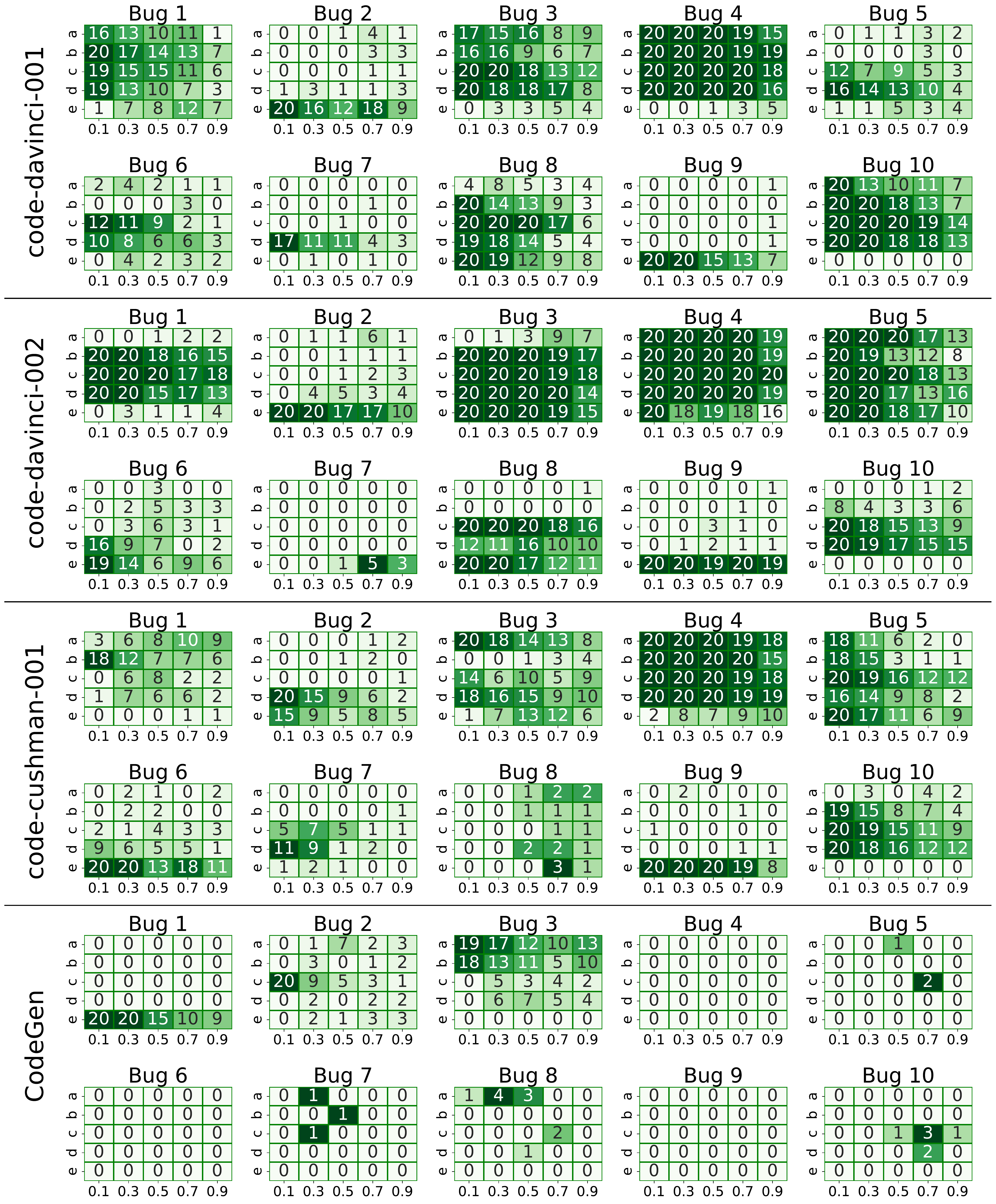}
    \caption{Results for all 4 LLMs represented as heatmaps. The maximum value for each small box is 20. A higher value indicates more success by LLM in generating repair and is highlighted with a darker shade. All bugs were repaired at least once by at least one LLM.}
    \label{fig:results-combined}
\end{figure*}

We set up our experimental framework for each LLM, generating 20 responses for every combination of bug, temperature, and instruction variation. The responses are counted as successful repairs if they pass functional and security tests. The number of successful repairs is shown as heat-maps in Figure~\ref{fig:results-combined}. The maximum value for each element is 20, i.e., when all responses were successful repairs.





%
\subsection{RQ1: Can LLMs fix hardware security bugs?}
\begin{figure}[]
    \includegraphics[width=0.9\linewidth]{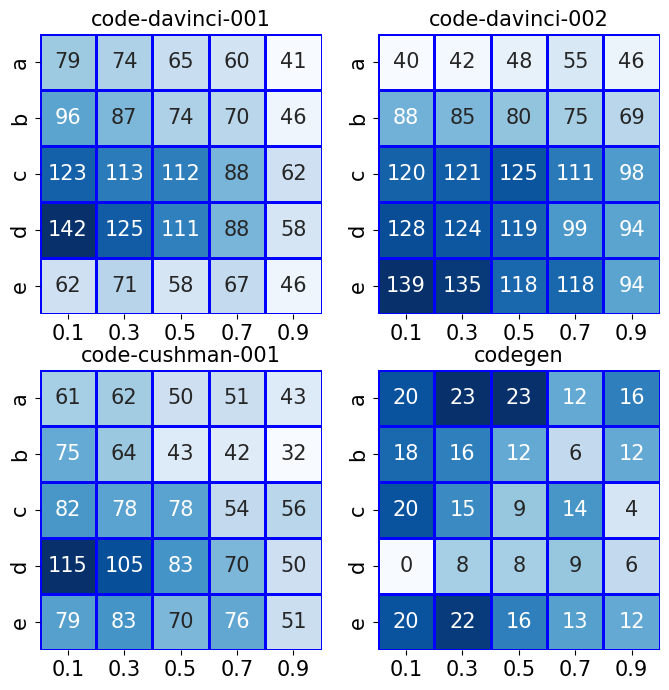}
    \caption{Results showing the performance of each LLM across all bugs in the form of heatmaps. Each small square shows the number of correct repairs for the corresponding instruction variation and temperature of the LLM. The maximum possible value is 200. A higher value indicates more success in generating repairs and is shaded in a darker color.}
    \label{fig:results-across-bugs}
\end{figure}

This work shows that LLMs can repair simple hardware bugs. \texttt{code-davinci-001}, \texttt{code-davinci-002}, and \texttt{code-cushman-001} yielded at least one successful repair for every bug in our dataset. \texttt{CodeGen} was successful for 7 out of 10 bugs.
In total, we requested 20,000 repairs out of which 6376 were correct, a success rate of 31.9\%.  The key here lies in selecting the best-observed parameters for each LLM. \texttt{code-davinci-001} performs best at variation $d$, $temp$ 0.1 producing 71\% correct repairs. \texttt{code-davinci-002}, \texttt{code-cushman-001} and \texttt{CodeGen} perform best at $(e,0.1)$, $(d,0.1)$ and $(a, 0.3$ and $0.5)$ with success rates of 70\%, 58\% and 12\% respectively. Performance of these LLMs across bugs is shown in \autoref{fig:results-across-bugs}.

We can fine-tune the parameters for each bug. The choice of the right combination of model, instruction variations and temperature can yield near-perfect results. We present these best-observed settings in Table \ref{tbl:optimal-settings}. Under these settings, Bug 7 has a success rate of 85\% and the rest have a success rate of 100\%.

\begin{table}[h]
\caption{Best-observed settings for each bug. `dv1', `dv2' and `cus' stand for \texttt{code-davinci-001}, \texttt{code-davinci-002} and \texttt{code-cushman-001}. Within the settings arrays, the first element is the LLM, the second is a set of instruction variations and the third is a set of temperatures.}
\label{tbl:optimal-settings}
\begin{tabular}{cl}
\hline
Bug                   & Best Setting                                                                                                                                                                                                                                                    \\ \hline\hline
1                     & {[}dv1,b,0.1{]} {[}dv2,(b,d),(0.1,0.3){]} {[}dv2,c,(0.1,0.3,0.5){]}                                                                                                                                                                                                \\ \hline
2                     & {[}cus,d,0.1{]} {[}dv1,e,0.1{]} {[}dv2,e,(0.1,0.3){]}                                                                                                                                                                                                              \\ \hline
3                     & \begin{tabular}[c]{@{}l@{}}{[}cus,a,0.1{]} {[}dv1,c,(0.1,0.3){]} {[}dv1,d,0.1{]}\\ {[}dv2,(b,c,e),(0.1,0.3,0.5){]} {[}dv2,d,(0.1,0.3,0.5,0.7){]}\end{tabular}                                                                                                      \\ \hline
4                     & \begin{tabular}[c]{@{}l@{}}{[}cus,(a,c,d),(0.1,0.3,0.5){]} {[}cus,b,(0.1,0.3,0.5,0.7){]}\\ {[}dv1,(a,b),(0.1,0.3,0.5){]} {[}dv1,(c,d),(0.1,0.3,0.5,0.7){]}\\ {[}dv2,(a,b,d),(0.1,0.3,0.5,0.7){]} \\ {[}dv2,c,(0.1,0.3,0.5,0.7,0.9){]} {[}dv2,e,0.1{]}\end{tabular} \\ \hline
5                     & \begin{tabular}[c]{@{}l@{}}{[}cus,(c,e),0.1{]} {[}dv2, (a,c),(0.1,0.3,0.5){]} {[}dv2,b,0.1{]}\\ {[}dv2, (d,e),(0.1,0.3){]}\end{tabular}                                                                                                                            \\ \hline
6                     & {[}cus,e,(0.1,0.3){]}                                                                                                                                                                                                                                              \\ \hline
7                     & {[}dv1,d,0.1{]}                                                                                                                                                                                                                                                    \\ \hline
8                     & \begin{tabular}[c]{@{}l@{}}{[}dv1,(b,e),0.1{]} {[}dv1,c,(0.1,0.3,0.5){]}\\ {[}dv2,c,(0.1,0.3,0.5){]} {[}dv2,e,(0.1,0.3){]}\end{tabular}                                                                                                                            \\ \hline
9   & \begin{tabular}[c]{@{}l@{}}{[}cus,e,(0.1,0.3,0.5){]} {[}dv1,e,(0.1,0.3){]} \\ {[}dv2,e,(0.1,0.3,0.7){]}\end{tabular}                                                                                                                                               \\ \hline
10                    & \begin{tabular}[c]{@{}l@{}}{[}cus,(c,d),0.1{]} {[}dv1,a,0.1{]} {[}dv1,(b,d),(0.1,0.3){]} \\ {[}dv1,c,(0.1,0.3,0.5){]} {[}dv2,(c,d),0.1{]}\end{tabular}                                                                                                             \\ \hline
\end{tabular}
\end{table}

\subsection{RQ2: What bugs are amenable to repair?}

\begin{figure}[]
    \includegraphics[width=\linewidth]{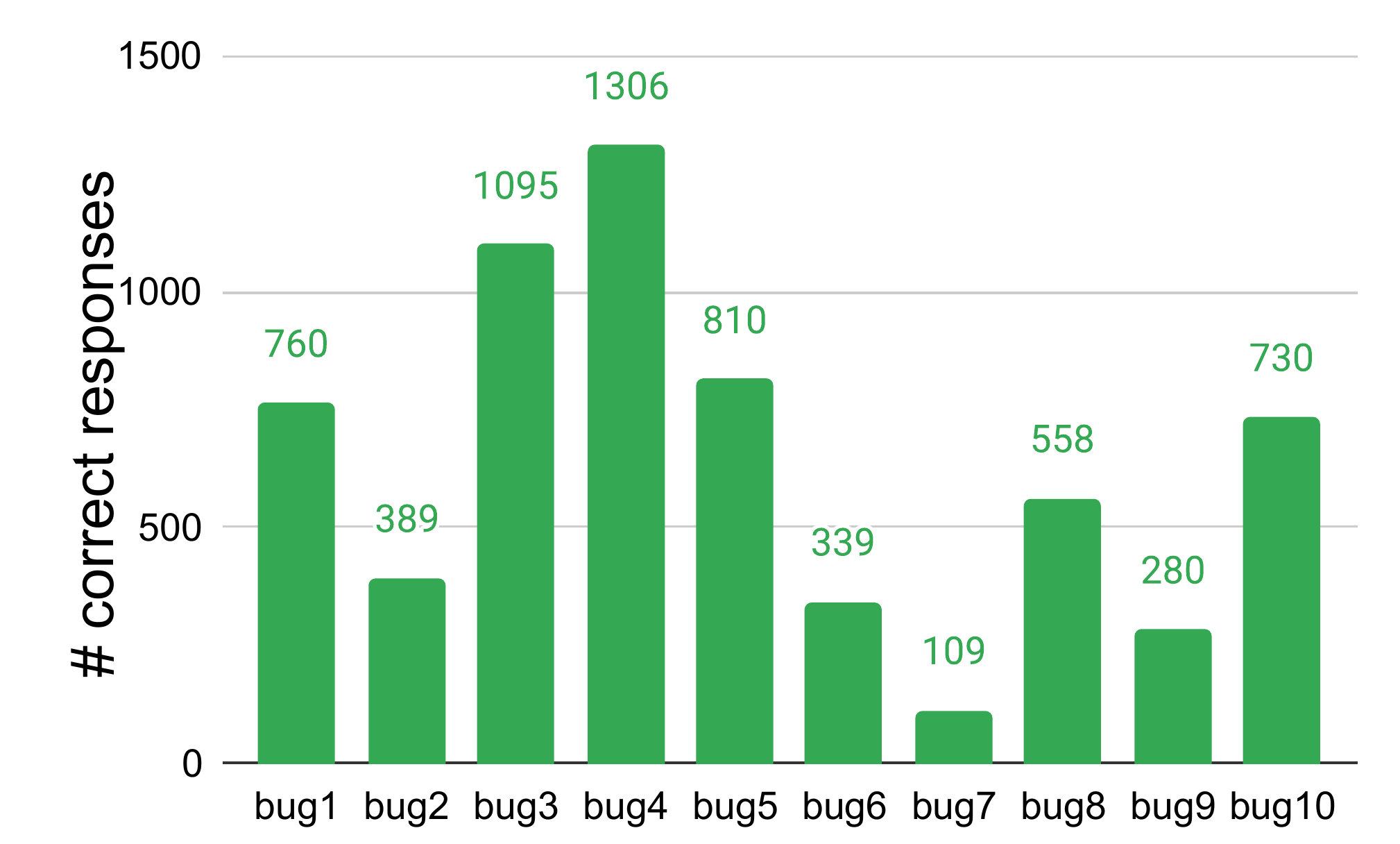}
    \caption{Number of correct repairs per bug. The number above each bar shows the sum of successful repairs across all LLMs for the corresponding bug. The maximum possible value is 2000. A higher value indicates that the bug was repaired more times.}
    \label{fig:results_bugs}
\end{figure}

The cumulative number of correct repairs for each bug is shown in \autoref{fig:results_bugs}.
Bugs 3 and 4 were the best candidates for repair with success rates of over 50\%. These are examples from MITRE where the signal names used clearly indicate their intended purposes. For the \textbf{Grant Access} module, the signals of concern are \texttt{grant\_access} and \texttt{usr\_id} used in successive lines.  LLMs are able to interpret the intended functionality that the \texttt{usr\_id} should be compared before granting access. Most successful repairs either flipped the order of blocking assignments or lumped them into an assignment using the ternary operator. Similarly, \textbf{Trustzone Peripheral} uses  signal names \texttt{data\_in\_security\_level} and \texttt{rdata\_security\_level} which illustrate their intended functionality.

Bugs 2, 6, 7, and 9 were the hardest to repair with success rates of under 25\%. Bugs 2 and 9 had the same bug of a register holding security settings not initialized under reset. This was difficult to repair because a fix required the creation of an always block with an appropriate reset as well as re-creating the previous intended functionality. Bug 7 was the hardest to repair because it was the only bug that required a line to be removed without replacement as a fix. We hypothesize that Bug 6 was hard to fix because it was difficult to phrase the fix instruction according to the description of the bug provided by Opentitan. The fix relies on asserting the fsm alert signal when escalation signal is high, but this condition is represented in code as \texttt{if (escalate\_en\_i != lc\_ctrl\_pkg::Off) } which is harder to grasp by the LLMs.

\subsection{RQ3: How should prompts be engineered to assist the repair of hardware bugs?}

The 5 variations from $a$ to $e$ increase in the level of detail. In general, apart from CodeGen, all LLMs do better with more detail being provided to assist repair as shown in \autoref{fig:trend-across-models}(a).
Variations $c$-$e$ perform better than variations $a$ and $b$. They include a fix instruction after the buggy code in comments, giving credence to the use of two separate instructions per prompt (one before and one after the bug in comments).
Variation $d$ has the highest success rate among OpenAI LLMs and is therefore our recommendation for bug fixes. The use of a fix instruction in `pseudo-code' fashion leads to the best results.
There is variation within LLMs for the best-observed instruction variation, e.g., \texttt{code-davinci-002} and \texttt{CodeGen} perform best at $e$.

\subsection{RQ4: Does the temperature matter?}
\begin{figure}[]
    \includegraphics[width=\linewidth]{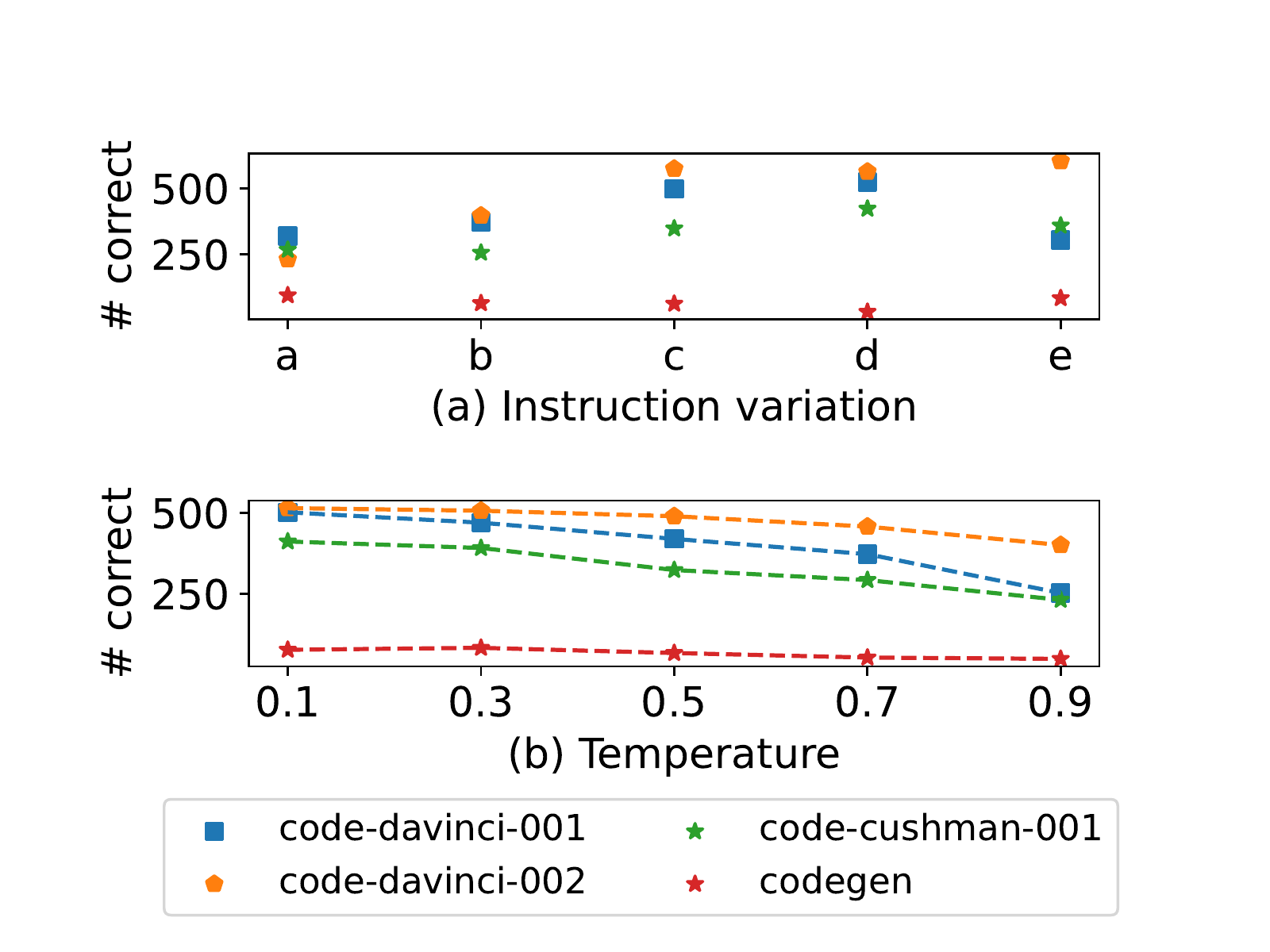}
    \caption{Results: Trends Across Models. The top graph shows the number of correct repairs for LLMs for specified instruction variations. The bottom graph shows the number of correct repairs for LLMs for specified temperature. The maximum value for each data point is 1000.}
    \label{fig:trend-across-models}
\end{figure}

A higher temperature allows the LLM to be more creative in its responses. As shown in \autoref{fig:trend-across-models}(b), the LLMs perform better at lower temperatures. All OpenAI LLMs perform best at $t=0.1$ and CodeGen performs best at $0.3$. A lower temperature leads to less variation in responses as well, implying that the less creative responses are more likely to be correct repairs.

\subsection{RQ5: Are some LLMs better than others?}
The \texttt{code-davinci-002} LLM was the best performing, producing 2371 correct repairs out of 5000, giving it a success rate of 47.4\%. \texttt{code-davinci-001}, \texttt{code-cushman-001} and \texttt{CodeGen} had success rates of 40.4\%, 33.1\% and 6.68\% respectively. The large difference between OpenAI LLMss and \texttt{CodeGen} is caused by \texttt{CodeGen} being a much smaller LLM, having 16 billion parameters compared to the OpenAI LLMs that are based on GPT-3's $\sim$175B parameters (the exact number of parameters for each of the OpenAI LLMs are not public).
Additionally, \texttt{code-cushman-001} is slightly 
inferior to the \texttt{davinci} LLMs because it was designed to be quicker. This may mean that it has fewer parameters or that it has been trained over less data or both.

\subsection{Comparison with CirFix}
\label{subsec:results-cirfix}

A comparison of our work with CirFix is shown in Table \ref{tbl:cirfix-comparison-results}. We use the best-performing LLM (\texttt{code-davinci-002}) at $t=0.1$ and generate one repair each for variations $a$ and $b$. This is done to closely mirror the use case of CirFix. By comparing the first example produced by the LLM, we evaluate only one attempt at repair. This attempt is manually evaluated for correctness. We use variation a for the primary comparison as this variation uses no instructions to assist repairs. This variation produces 17.5 correct repairs as compared to CirFix's 16. The half repair corresponds to fixing one numeric error out of 2 for the first benchmark. To elicit the power of LLMs, we use variation b which includes a description of the type of bug to assist repair. We use the brief descriptions of bugs provided in CirFix's GitHub repository. Variation $b$ fixes 20 of the 32 benchmarks and collectively, LLMs (both variations) are able to repair 23.5 of the bugs.


\begin{table}[]
\caption{Comparison on CirFix benchmarks. A successful repair is shown as y. We use two instruction variations for this comparison. An element - | y means that the repair using variation $a$ was not successful but using variation $b$ was. The element 1/2 means that 2 errors were used in the description of a single fault/bug and 1 out of the 2 was successfully repaired. }
\label{tbl:cirfix-comparison-results}
\resizebox{\linewidth}{!}{
\begin{tabular}{llcc}
\hline
\textbf{Project}                                                & \textbf{Defect Description}                                                                                                      & \textbf{CirFix}           & \textbf{\begin{tabular}[c]{@{}c@{}}LLM\\ var $a$ | $b$\end{tabular}} \\ \hline
decoder                                                         & Two numeric errors                                                                                                               & y                         & 1/2 | 1/2                                                        \\
(3 to 8)                                                        & \cellcolor[HTML]{C0C0C0}Incorrect assignment                                                                                     & \cellcolor[HTML]{C0C0C0}- & \cellcolor[HTML]{C0C0C0}- | -                                    \\ \hline
first\_counter                                                  & Incorrect sensitivity list                                                                                                       & y                         & y | y                                                            \\
overflow                                                        & \cellcolor[HTML]{C0C0C0}Incorrect increment of counter                                                                           & \cellcolor[HTML]{C0C0C0}y & \cellcolor[HTML]{C0C0C0}- | y                                    \\
                                                                & Incorrect reset                                                                                                                  & y                         & y | y                                                            \\ \hline
flip\_flop                                                      & Incorrect conditional                                                                                                            & y                         & - | -                                                            \\
                                                                & \cellcolor[HTML]{C0C0C0}if-statement branches swapped                                                                            & \cellcolor[HTML]{C0C0C0}y & \cellcolor[HTML]{C0C0C0}| y                                      \\ \hline
fsm\_full                                                       & Incorrect case statement                                                                                                         & -                         & - | -                                                            \\
                                                                & \cellcolor[HTML]{C0C0C0}\begin{tabular}[c]{@{}l@{}}Assignment to next state and\\ default in case statement omitted\end{tabular} & \cellcolor[HTML]{C0C0C0}y & \cellcolor[HTML]{C0C0C0}y | y                                    \\
                                                                & State omitted from senslist                                                                                                      & -                         & - | y                                                            \\
                                                                & \cellcolor[HTML]{C0C0C0}Incorrect blocking assignments                                                                           & \cellcolor[HTML]{C0C0C0}- & \cellcolor[HTML]{C0C0C0}- | -                                    \\ \hline
lshift\_reg                                                     & Incorrect blocking assignments                                                                                                   & y                         & - | y                                                            \\
                                                                & \cellcolor[HTML]{C0C0C0}Incorrect conditional                                                                                    & \cellcolor[HTML]{C0C0C0}y & \cellcolor[HTML]{C0C0C0}- | y                                    \\
                                                                & Incorrect sensitivity list                                                                                                       & y                         & y | y                                                            \\ \hline
mux\_4\_1                                                       & Three numeric errors                                                                                                             & -                         & y | y                                                            \\
                                                                & \cellcolor[HTML]{C0C0C0}Hex instead of binary numbers                                                                            & \cellcolor[HTML]{C0C0C0}- & \cellcolor[HTML]{C0C0C0}y | y                                    \\
                                                                & 1 bit instead of 4 bit output                                                                                                    & -                         & y | y                                                            \\ \hline
i2c                                                             & Incorrect sensitivity list                                                                                                       & y                         & - | -                                                            \\
                                                                & \cellcolor[HTML]{C0C0C0}Incorrect address assignment                                                                             & \cellcolor[HTML]{C0C0C0}- & \cellcolor[HTML]{C0C0C0}y | -                                    \\
                                                                & No command acknowledgement                                                                                                       & y                         & y | y                                                            \\ \hline
sha3                                                            & Off-by-one error in loop                                                                                                         & y                         & y | -                                                            \\
                                                                & \cellcolor[HTML]{C0C0C0}Incorrect assignment to wire                                                                             & \cellcolor[HTML]{C0C0C0}- & \cellcolor[HTML]{C0C0C0}y | -                                    \\
                                                                & Skipped buffer overflow check                                                                                                    & y                         & - | -                                                            \\
                                                                & \cellcolor[HTML]{C0C0C0}Incorrect bitwise negation                                                                               & \cellcolor[HTML]{C0C0C0}- & \cellcolor[HTML]{C0C0C0}y | y                                    \\ \hline
tate\_pairing                                                   & Incorrect logic for bitshifting                                                                                                  & -                         & - | -                                                            \\
                                                                & \cellcolor[HTML]{C0C0C0}Incorrect instantiation of modules                                                                       & \cellcolor[HTML]{C0C0C0}- & \cellcolor[HTML]{C0C0C0}y | y                                    \\
                                                                & Incorrect operator for bitshifting                                                                                               & -                         & y | y                                                            \\ \hline
\begin{tabular}[c]{@{}l@{}}reed\_solomon\\ decoder\end{tabular} & \begin{tabular}[c]{@{}l@{}}Insufficient register size\\ for decimal values\end{tabular}                                          & -                         & - | -                                                            \\
                                                                & \cellcolor[HTML]{C0C0C0}Incorrect sensitivity list for reset                                                                     & \cellcolor[HTML]{C0C0C0}y & \cellcolor[HTML]{C0C0C0}y | y                                    \\ \hline
\begin{tabular}[c]{@{}l@{}}sdram\\ controller\end{tabular}      & \begin{tabular}[c]{@{}l@{}}Incorrect assignments to registers\\ during synchronous reset\end{tabular}                            & y                         & y | 1/2                                                          \\
                                                                & \cellcolor[HTML]{C0C0C0}Numeric error in definitions                                                                             & \cellcolor[HTML]{C0C0C0}- & \cellcolor[HTML]{C0C0C0}y | y                                    \\
                                                                & Incorrect case statement                                                                                                         & -                         & - | y                                                            \\ \hline
                                                                & \multicolumn{1}{c}{}                                                                                                             & 16                        & 17.5 | 20                                                          \\ \cline{3-4} 
\end{tabular}
}
\end{table}

\section{Discussion and Limitations\label{sec:dis}}

Our results show that LLMs have a lot of potential for bug repair. At the present, some assistance is required from the designer to identify the location and nature of the bug. This may be overcome by using other tools to localize the bugs and better design practices such as comments explaining the functionality of the design. Currently, the designer may also be needed to pick between a few options produced by the LLMs. This is where static analysis tools like CWEAT and other bug detection tools may come in to complete the loop by suggesting a repair that is correct to a high degree of confidence.

A limitation of the work is the subjectivity of instruction variations. Although the Bug instructions devised are inspired by the descriptions in CWEs, the Fix instructions are devised according to the knowledge and experience of the authors. Our work reveals the importance of these variations as subtle changes can affect the response generated by LLMs. Devising 5 categories is an attempt to standardize this process, but more varieties are probably needed to study their effects better. 
Moreover, instructions are challenging to generalize across different bugs. Ideally, a designer would want variation \textbf{a} to fix all bugs because no instructions are needed. But since more information is needed according to the particular instance of the bug for a higher probability of a successful fix, it is a challenge to form a small set of instructions, e.g., if LLMs are able to produce a successful fix with the Bug Instruction "Improper FSM" instead of "FSM has an unreachable state", that would be better.

Another limitation of the current framework is that the functional and security evaluations are not exhaustive. Security evaluation is dependent on the security objectives for the design and can not truly be exhaustive~\cite{aftabjahani_special_2021}. With this in mind, we limit the security evaluation to the particular bug that makes the design insecure. Ideally, efforts should be made to check that a fix does not result in another kind of bug. 
Functional evaluation is needed because a design that is secure but not functional is useless. For the CWE examples, we were able to build exhaustive testbenches because the designs were low in complexity and had only one or two modules. Ideally, the functional testbenches should be exhaustive for other examples too. But this would be very time-consuming as the size of the designs gets very large. It would be a difficult task to write testbenches for these complex SoCs and simulating the designs according to the software provided by OpenTitan and Hack@DAC takes a lot of time, e.g., design verification of OpenTitan examples takes $\sim$10 minutes and it takes $\sim$15 minutes to simulate the Hack@DAC-21 SoC.
Therefore, we chose to build custom testbenches that test the code a repair could impact. 

The use of end-tokens is another area of subjectivity that influences the success rate of repairs. Some strategies are intuitive like using the end line token as an end token for a bug that is present in only one line. Others may require more creativity because some lines of code can be written in multiple ways. A repair that spans multiple conditional statements, e.g., 
\begin{verbatim}
    if (~resetn) begin locked <= 0; end 
    else if(unlock) begin locked <= d; end 
    else begin locked <= locked; end 
\end{verbatim}
may not be completely produced if the keyword \textbf{end} is used as a stop token.
On the other hand, not limiting a response with an appropriate stop token may mean that the LLM produces the correct repair but then adds more code that contains a syntax error or affects functionality. 
We use a post-processing script to minimize syntax errors. This involves adding/removing the \texttt{end} keyword as needed. When the LLM generates a repair, that repair is a substitute for the bug only. The number of \texttt{begin} and \texttt{end} keywords are counted. If the numbers are same, nothing is to be done, and the repair is inserted in place of the bug. If the number of begins are greater by an amount $n$, \texttt{end} is added at the end of the repair $n$ times.
If the number of ends are greater by an amount $n$, the first $n$ instances of \texttt{end} are removed.

The LLMs are very quick in generating repairs. The 20 responses per request are generated in under a minute. While trying to find a repair for a bug, a Verilog designer should have enough suggested repairs very quickly. The designer can then choose the best suggestion as the repair. In our experiments, we faced some challenge because of token limits set by the OpenAI API. Since we were generating thousands of requests with a limited number of token keys, we had to wait for a minute ever time we reached the limit. This raised our generation of repair time to $\sim$20 minutes per LLM.

To evaluate security-related hardware bugs, a large number of benchmarks are needed that show these defects in designs. Our work takes a step in this direction. We believe more examples are needed to make more conclusive claims about repair techniques.



\section{Conclusions and Future Work\label{sec:conc}\label{sec:conclusions}}

By choosing the right parameters and providing the right prompts, LLMs can fix the hardware bugs in our corpus. All the bugs had at least one successful repair and 9 of the 10 had 100\% correct responses with the best set of parameters. 
We have found that in instances where signal names and comments implicate the functionality, LLMs have a high success rate. Conversely, fixes that span more than 1 line or require the removal of a buggy line are harder to repair.
Detailed instructions to assist repair tend to achieve higher success rates with variation $d$ using a Fix instruction that uses pseudo-code-like language performing the best. LLMs at lower temperatures and bigger LLMs perform better than LLMs at higher temperatures and LLMs with fewer parameters. LLMs do a better job at fixing function-related bugs in Verilog relative to the program repair mechanism in CirFix. We propose the following directions for future work:
\begin{itemize}
    \item Test a hybrid approach for security-related bugs. Use Linters, Formal Verification, fuzzing, fault localization, and static analysis tools for detection and LLMs, oracle-guided modifying algorithms for repair. An ensemble of these options is likely to have more success than one technique alone.
    \item Fine-tune LLMs over HDLs and see if the performance improves. This improves the generation of functional code \cite{thakur_benchmarking_2022}.
    \item Explore the repair of functional bugs using LLMs with the full sweep of parameters. We only used one set of parameters that performed the best in our experiments.
    \item Build a database of security-related hardware bugs. Ongoing efforts like Trusthub's Vulnerability Database \cite{noauthor_trust-huborg_2023} can be consolidated with our examples to build standard benchmarks. 
\end{itemize}


\bibliographystyle{ACM-Reference-Format}

\bibliography{lit/hw-repair-llm}

\section*{Appendix\label{sec:appendix}}


\subsection*{Compute environment}
All experiments were conducted on a Intel Core i5-10400T CPU @2GHzx12 processor with 16 GB RAM. Operating system Ubuntu 20.04.5 LTS was used.

\subsection*{Open source details}

There are a few parts of our experimental framework where we could not provide fully open-source access:

\begin{itemize}
\item Verific: We used Verific libraries provided by Verific under an academic license. Please contact Verific to get access to their products.

\item CWEAT: We requested CWEAT code from the authors of the paper "Don't CWEAT It: Toward CWE Analysis Techniques in Early Stages of Hardware Design" \cite{ahmad_dont_2022}. The paper is available at https://dl.acm.org/doi/abs/10.1145/3508352.3549369. Please contact the authors for use/help with their codebase.

\item CirFix: We used the CirFix benchmarks and results provided in the open-source github repository provided by the authors of the paper ``CirFix: automatically repairing defects in hardware design code.'' ~\cite{ahmad_cirfix_2022}
\url{https://github.com/hammad-a/verilog_repair}. Please contact the authors about use of their tools. Their paper is available at\\
https://dl.acm.org/doi/10.1145/3503222.3507763.

\item Hack@DAC SoC: We use the SoC used during the 2021 competition. Please contact them at info@hackatevent.org for more information/access.
\end{itemize}

\end{document}